**Title: Transition from turbulent to coherent flows in confined three-dimensional active fluids**

**Authors:** Kun-Ta Wu[1], Jean Bernard Hishamunda[1], Daniel T.N. Chen[1], Stephen J. DeCamp[1], Ya-Wen Chang[2], Alberto Fernández-Nieves[2], Seth Fraden[1]* and Zvonimir Dogic[1]*

**Affiliations:**

[1]Department of Physics, Brandeis University, 415 South Street, Waltham, Massachusetts 02453, USA.

[2]School of Physics, Georgia Institute of Technology, 837 State Street, Atlanta, Georgia 30339, USA.

*Correspondence to: fraden@brandeis.edu, zdogic@brandeis.edu

**Abstract**: Transport of fluid through a pipe is essential for the operation of macroscale machines and microfluidic devices. Conventional fluids only flow in response to external pressure. We demonstrate that an active isotropic fluid, comprised of microtubules and molecular motors, autonomously flows through meter-long three-dimensional channels. We establish control over the magnitude, velocity profile and direction of the self-organized flows, and correlate these to the structure of the extensile microtubule bundles. The inherently three-dimensional transition from bulk-turbulent to confined-coherent flows occurs concomitantly with a transition in the bundle orientational order near the surface, and is controlled by a scale-invariant criterion related to the channel profile. The non-equilibrium transition of confined isotropic active fluids can be used to engineer self-organized soft machines.

**One Sentence Summary:** An isotropic fluid composed of nano-sized motors organizes into an autonomous machine that pumps fluid through long channels.

**Main Text:** Conventional non-equilibrium pattern forming systems, such as Rayleigh-Benard convection, are powered by continuous injection of energy through macroscopic external boundaries (*1*). In comparison, hierarchically organized living matter is driven away from equilibrium by the motion of microscopic energy-transducing animate constituents. Starting with molecular motors and continuing up to subcellular organelles, cells, tissues and entire organisms, each biological element is constructed from energy-consuming components. At each level of hierarchy these animate elements push and pull on each other, thus generating local active stresses. Collectively these stresses enable diverse functionalities that are essential for survival of living organisms, such as cell division and motility (*2-8*). Under geometrical confinements active building blocks such as motile cells, tissues and organisms can also organize into coherently moving non-equilibrium states (*9-17*). Creating active matter systems from the bottom-up that mimic the remarkable properties of the living matter remains a fundamental challenge (*18-20*). Recent studies have revealed emergence of diverse complex patterns in synthetic systems of active matter (*21-24*). The next step is to elucidate conditions that transform chaotic dynamics of these systems into coherent long-ranged motion that can be used to harvest energy and thus power various micromachines (*25-29*).

Here, we study 3D active fluids and demonstrate an essential difference with their conventional counterparts. The Navier-Stokes equations dictate that a conventional fluid comprised of inanimate constituents will flow only in response to an externally imposed body force, or stress and pressure gradients (29). This is no longer true for active fluids. Indeed, in living organisms, the entire cellular interior can assume large-scale coherent flows in absence of any externally imposed stresses, a phenomenon known as cytoplasmic streaming (30-32). Despite recent advances using living bacterial suspensions (13, 14, 33, 34), creating tunable synthetic active fluids that exhibit autonomous long-ranged flows on length scales large compared to constituent units remains a challenge. We use a 3D microtubule-based isotropic active fluid whose bulk turbulent flows are driven by continuous injection of energy through the linear motion of the constituent kinesin motors (24, 35). We find that confinement robustly transforms locally-turbulent dynamics of such active fluids into globally-coherent flows that persist on meter scales. Our experiments demonstrate that non-equilibrium transitions of synthetic active materials can be used to engineer self-organized machines in which nanometer sized molecular motors collectively propel fluid on macroscopic scales.

**Microtubule-based active isotropic fluids:** The active fluid we study is comprised of microtubule filaments, kinesin motor clusters and depleting polymer (Fig. 1A) (*24, 35*). Kinesin motors are bound into synthetic clusters with tetrameric streptavidin (*36, 37*). The depleting polymer induces microtubule bundling (*38*), while the kinesin clusters simultaneously bind to and move along the neighboring filaments. For aligned microtubules with opposite polarity, kinesin clusters generate interfilament sliding that drives the flow of the background fluid (Fig. 1A) (*39*). At finite concentrations (≥ 0.5 mg/mL) active microtubule bundles form a self-rearranging 3D isotropic network, whose dynamics is comprised of repeating cycles of bundles extending, buckling, fracturing and annealing. Such dynamics effectively drives the flow of the background fluid that is mostly comprised of an aqueous buffer (~99.95%). In bulk suspensions, microtubule based active networks generate turbulent-like flows that exhibit neither long-range order nor net material transport. The structure of these flows and the associated vorticity fields are visualized by doping active fluids with fluorescent tracer particles (Fig. 1D, movie 1). Such analyses reveal a characteristic vortex size of ~100 μm with an average lifetime of ~3 sec.

Energy-efficient kinesin motors allow for assembly of bulk 3D active fluids that maintain steady-state turbulent flows for up to ten hours. This unique feature allows us to quantify the time-averaged of various steady flows. Furthermore, filamentous microtubules effectively couple to the background fluid. Therefore, even at volume fractions as low as 0.05%, the microtubule network drives the bulk turbulent flows. At such low concentrations, the network has very large pore size (~10 μm), thus its structure can be visualized and quantitatively analyzed with optical microscopy. This property allows us to correlate the properties of the emergent flows to the structure of the underlying active network.

**Confinement induced coherent flows:** We found that 3D toroidal confinements can effectively transform directionless turbulent flows of bulk fluids into coherent circular currents capable of transporting materials over macroscopic scales (Fig. 1B, 1E). Similar phenomena were also observed for cylindrical confinements, demonstrating that toroidal geometry is not an essential requirement for formation of coherent flows (Fig. 1F). The autonomous circular currents persisted for hours, and ceased only after the available chemical energy (ATP) is depleted (Fig. 1C). The flow coherence was quantified by the order parameter $\Phi \equiv \langle v_\theta / |\mathbf{v}| \rangle$, where $v_\theta$ is the flow velocity along the azimuthal direction ($\hat{\boldsymbol{\theta}}$), and $|\mathbf{v}|$ is the flow speed. $\Phi=1$ indicates perfect circular flows, while $\Phi=0$ indicated no net transport along the tangential direction. We measured $\Phi$~0.6, and the magnitude of the order parameter remained fairly constant over the entire sample lifetime (Fig. 1C).

Not all geometries support coherent flows. For example, reducing the height of a toroid or a cylinder below a critical value completely suppressed coherent flows and yielded turbulent-like dynamics (Fig. S1, movie 2). To determine geometries that support self-organized circular flows, we varied the widths, heights and radii of toroids with rectangular cross-sections, as well as radii and heights of the cylindrical geometries (Figs. 2A). From an extensive data set we draw the following observations: First, self-organized flows are robust, occurring over a range of confining geometries whose volumes span more than four orders of magnitude (from 0.006 to 18 µL). This range could be larger, since we have determined neither the upper nor the lower volume that supports circular flows. Second, the self-organized circular flows are equally likely to be left- or right-handed. Third, for both toroids and cylinders, coherent flows disappear as the confining geometry becomes either too wide or too tall, or more precisely, when the ratio of the long to short side of the channel cross-section becomes greater than about 3.

To quantify this observation we plotted the circulation order parameter $\Phi$ as a function of the aspect ratio of toroids or cylinders: $\alpha \equiv (h-w)/\text{Max}(h,w)$, where $h$ and $w$ are the height and width of either a cylinder or toroid, and $\text{Max}(h,w)$ represents their larger value (Fig. 2B). Geometries with square-like profiles ($\alpha=0$) support robust flows. As the geometry becomes wider and/or shorter ($\alpha<0$), the circulating flows only formed for cross-sections that were sufficiently symmetric ($-0.6<\alpha<0$), otherwise no net flows were observed. In the complementary regime of tall and narrow confinements ($\alpha>0$) we observed a second transition from coherent to turbulent flow as the confinement geometry became taller or narrower ($\alpha>0.6$). This observation demonstrates the equivalency between the height and width (or vorticity and gradient direction) of the confining geometry.

We observed flows over a well-defined range of the parameter $\alpha$ ($-0.6<\alpha<0.6$). This criterion appears universal as it holds for both micron- and millimeter-scale confinements as well as for toroidal and cylindrical geometries. This data suggest that the transition to coherent flows is not controlled by the intrinsic lengthscale of the bulk fluid, which is the vortex size of 100 μm (Fig. 1D), but rather by the dimensionless aspect ratio, $\alpha$. For example, coherent circular flows are observed in cylinders with $w$ = 3.3 mm, which is 30 times larger than the vortex size of the bulk fluid. This is contrary to previous studies which concluded the need to confine flows smaller than the intrinsic lengthscale of the bulk system *(13, 34)*. The criterion we uncovered demonstrates that the transition to coherent flows is an intrinsic 3D phenomenon that cannot, even in principle, exist in 2D systems. As the channel height is reduced to the 2D limit ($h \to 0$), the confinement parameter reduces to a single value, $\alpha \to -1$, which is in the regime where coherent flows do not occur (Fig. 2B). In other words, the scale-invariant criterion does not exist in 2D because there is only one channel dimension ($w$) perpendicular to the flow, which has the effect of fixing the channel cross-section to a single value. In contrast, in 3D the cross-section has two dimensions ($h$, $w$), making it is possible to construct a dimensionless aspect ratio from the two dimensions of the channel cross-section that can continuously control the turbulent-to-coherent transition.

**Structure of the coherent flows:** To quantify the spatial structure of the self-organized circular flows, we measured time-averaged azimuthal velocity profiles $\langle v_\theta(r) \rangle$ in both toroidal and cylindrical geometries. For small toroidal widths ($w$ < 1,000 μm) coherent flows exhibited Poiseuille-like symmetric profiles, with a peak velocity at the channel midpoint and decreasing to zero at the boundary due to the no-slip condition (Fig. 2C). The peak velocity increased with channel height. For a 330 μm high and 650 μm wide toroid the maximum velocity was ~4 μm/sec. Increasing the height to 1,300 μm, for the same width, increased the peak velocity to ~10 μm/sec. For the largest toroids studied, peak velocities reached ~17 μm/sec, while for smallest micron-sized ones they were as small as ~1 μm/sec (Fig. 2C inset). The asymmetry of the time-averaged velocity profile increased with increasing toroid width. It became especially pronounced for cylindrical geometries where it increased from zero at the disk center, growing linearly to a maximum value before decreasing rapidly to zero at the outer edge (Fig. 2D). Such a distinguishing feature is universal for both micron- and millimeter-sized cylinders, indicating again that the flow structure is independent of the confinement size.

**Controlling handedness of coherent flows:** In mirror-symmetric toroids and cylinders coherent flows are equally likely to be either counter-clockwise (CCW) or clockwise (CW). To control the flow handedness we decorated the toroid outer edge with sawtooth notches, finding that a CCW geometry induces flows of same handedness and vice versa (Fig. 3A and movie 3). In contrast, decorating inner edges with the same pattern leads to the flows with opposite handedness (Fig. S2A-B). Coherent flows could be completely suppressed by decorating both inner and outer surface with the notches of same chirality (Fig. S2C-D). Decreasing the number of notches increased the circulation order parameter, $\Phi$, and even a single notch controlled the flow direction (Fig. 3D). Flow maps reveal that each notch creates a stationary vortex (Fig. 3B), and the back-flow associated with this vortex decreases the overall flow coherency (Fig. 3C).

We explored how the self-organized flow rates scale with the channel curvature and length, by either changing the toroid radii or stretching toroids into shapes resembling racetracks, while maintaining the same cross-section. The toroidal radius (curvature) does not affect the properties of the self-organized flows (movie 4 and Fig. S3). For racetrack geometries we varied the total circumference length from 14 to 48 mm (Fig. 4A and movie 5). Flow profiles in these geometries were nearly identical, demonstrating not only that the flow rate is independent of the pipe length, but also that the straight and curved segments of a racetrack provide equal driving power. To further confirm this finding, we designed a 20-cm long track (Fig. 4B). Coherent flow persisted in such a lengthy track; its rate remaining nearly the same as in a toroid of equivalent cross-section but shorter circumference (dashed curves in Fig. 4A).

**Coherent flows in complex confinements:** The self-organized coherent flows are remarkably robust self-organized patterns. To determine limits of their stability, we explored emergent dynamics in geometries of increasing complexities. First, coherent flows developed not only in simple toroids but also in a complex maze-like geometry comprised of a closed loop and many dead-end branches (Fig. 4C). Upon filling such geometry with active fluid, coherent flows developed in the closed loop, while remaining turbulent in the dead-end branches. However, unlike racetrack geometries, the flow rate was ~50% slower than in an equivalent toroid, indicating that the dead-end branches disturb the self-organized flows (dotted curves, and dashed black curve in Fig. 4A). Second, we have explored flows that emerge in two or three adjoining toroids, and found that such geometries support multiple self-organized states (Fig. 5 and movie 6)(*40*). For example, in one state, flows in the two adjoining toroids had opposite handedness, and were thus reinforced in the mutually adjoining region (Fig. 5A). In another state, the flow organized along the outer edge of the toroid doublet, therefore suppressing coherence in the adjoining region (Fig. 5B). Similarly, in triplet geometry we observed either flow circulating along the outer edge, leaving the center quiescent (Fig. 5C) or clockwise flow in one toroid and counterclockwise flow in the other two (Fig. 5D). Third, coherent flows also developed in confinements with sharp corners, such as a square-like geometry (Fig. 5E). However, sharp corners induced formation of stationary vortices, which generated back-flow, thus decreasing the circulation order parameter, when compared with a toroid with an equivalent cross-section. Finally, we found that the self-organized flows are not limited to microfluidic channels. For example, active fluids exhibited macroscopic flows both within a torus inscribed within a finite yield-stress elastomer (Fig. S4) (*41*), as well as a 1.1 meter-long plastic tube, (Fig. 6).

**Structure of the microtubule network:** To elucidate the microscopic mechanism that drives coherent flows we imaged the microtubule network structure of active fluids in both turbulent and coherent states and extracted the orientational distribution function (ODF) of the constituent bundle segments (Fig. 4A and movie 7) (*42*). In the channel center the measured ODFs were isotropic in both the coherent and turbulent states (indistinguishable from ODFs of bulk active fluids) (Fig. 4B). Emergence of the coherent flows was correlated with the change in the thickness of the nematic layer that wets the surfaces. We measured the spatial variation of the nematic order parameter in the direction perpendicular to the confining wall (Fig. 4C). In turbulent states the nematic order decays away from the boundaries with a characteristic lengthscale ($l \sim 30$ μm). Transition to coherent flows is accompanied with an increase in the thickness of the surface induced nematic layer ($l \sim 100$ μm). Similar to fluid flow profiles (Fig. 2C), lateral profiles of nematic order were asymmetric for circulating toroids with wide cross-section, while symmetric for narrow ones (Fig. 4C inset), implying that the shear rates generated by coherent flows drive the filament alignment. The angle between the confining surface and the self-organized nematic layer had a finite non-zero value of ~20 degrees (solid blue curve in Fig. 7B), suggesting that active stresses generated by the spatially distorted wall-bounded nematic layer push against the no-slip boundary and thus propel the fluid forward (*43*).

**Discussion:** We demonstrated that confining boundaries transform turbulent dynamics of bulk isotropic active fluids into long-ranged coherent flows. Our results suggest a mechanism by which this transition takes place. In conventional isotropic liquid crystals the bounding surface breaks the local symmetry and aligns the neighboring constituent particles (*44*). In active systems turbulent flows enhance this effect, magnifying and extending the wall-induced nematic order. We suggest that the nematic-wetting layer aligns to the surface at an oblique angle, thus generating an active surface stress that induces further flows. In turn, these flows enhance and stabilize the orientation of the wetting layer. This positive feedback loop yields a cooperative non-equilibrium transition to globally coherent flows, reminiscent of a phenomenon known as "backflow" in conventional liquid crystals, where mutually reinforcing coupling between the reorienting nematic director and the fluid velocity leads to emergent periodic structures (*45, 46*).

The boundary generated active stresses power coherent flows. This bears similarity to electroosmosis, in which an external electric field creates boundary layer stresses that lead to plug flow (*47*). The time-averaged coherent flows of active matter confined in cylinders resemble plug flows, in which the interior fluid away from the interface moves as a solid body. However, the velocity of electro-osmotic plug flow is independent of channel size, when the boundary layer is small; we observe that the coherent active flow velocity varies with channel size. Additionally, active flows in tori resemble Poiseuille flow, indicating pressure driven flow.

Emergence of coherent flows is determined by a universal scale-invariant criterion, which is related to the aspect ratio of the confining channel cross-section, rather than the overall size of the confining geometry. In particular, coherent flows take place when the channel cross-section is relatively symmetric: the width and height of the channel are within a factor of 3 of each other. Our findings are distinct from theoretical models that predict a transition from quiescent to coherent flows for 2D monodomain active liquid crystals, through an active analogue of the classical Fréedericksz transition (*25, 26*). First, our phenomenon is an inherently three-dimensional. The observation that the aspect ratio alone and not the absolute channel size controls the onset of coherent flows, forbids this transition from taking place in systems with lower dimensionality. Second, the coherent flows form in an active isotropic fluid, instead of a liquid crystalline one. Third, the flows also emerge from the turbulent state, while the theory predicts the formation of such flows from a quiescent, single domain nematic.

We demonstrate that non-equilibrium transitions in active matter can be used for the robust assembly of self-organized machines that produce useful work by powering large-scale fluid flow. These machines are fueled by the collective motion of the constituent nanometer sized motors, and thus symbols living organisms. However, it is not evident if the mechanism we describe has direct relevance to cell biology. Coherent cytoplasmic flows in living cells are powered by cytoskeletal filaments that are permanently affixed at the cellular cortex (30-32), while in our system the active elements are dispersed throughout the sample interior and dynamically self-organize at the solid boundaries that confine the fluid. Explaining the scale invariant criterion for the onset of coherent flows and their velocity profiles represents a challenge for active matter theory.

**Figure Captions:**

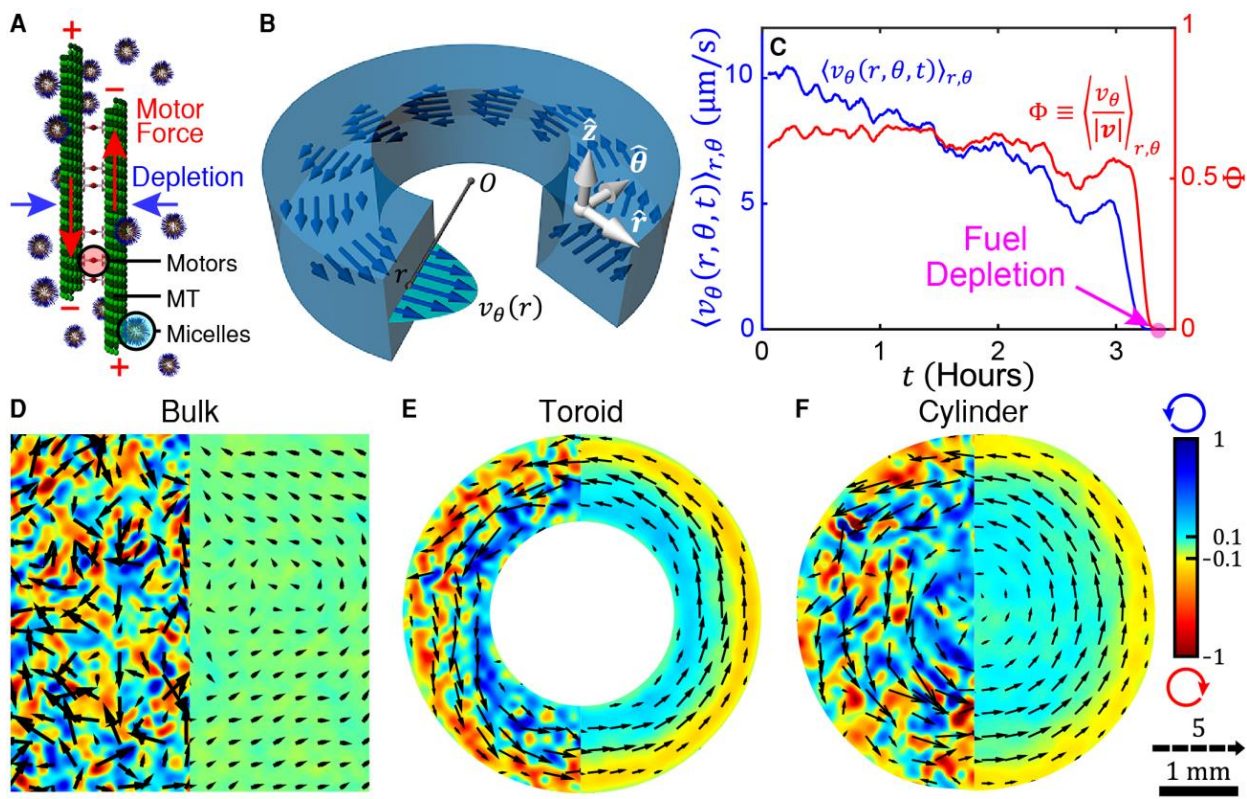

**Fig. 1**. **Coherent macroscopic flows of confined active fluids.** (**A**) The three main constituents of an active fluid: microtubules, kinesin clusters which power interfilament sliding, and pluronic micelles that act as a depleting agent. (**B**) Schematic of a coherent flow that emerges when active fluids are confined in a toroid. (**C**) Time evolution of the averaged azimuthal velocity (blue) and the circulation order parameter (red) demonstrates that circular flows cease upon ATP depletion. (**D**) A flow map demonstrates that bulk active fluids exhibit turbulent flows that do not transport material. The vector field indicates the local flow velocity normalized by the mean flow speed. The color map represents the normalized vorticity distribution with blue and red tones representing counter-clockwise and clockwise vorticities. Left and right halves are instant and time-averaged plots, respectively. (**E**) An active fluid in a toroidal confinement exhibits persistent circular flows. (**F**) Coherent flows in a cylindrical confinement. Confinement heights are 1.3 mm. D-F share the same scale, velocity and color bars.

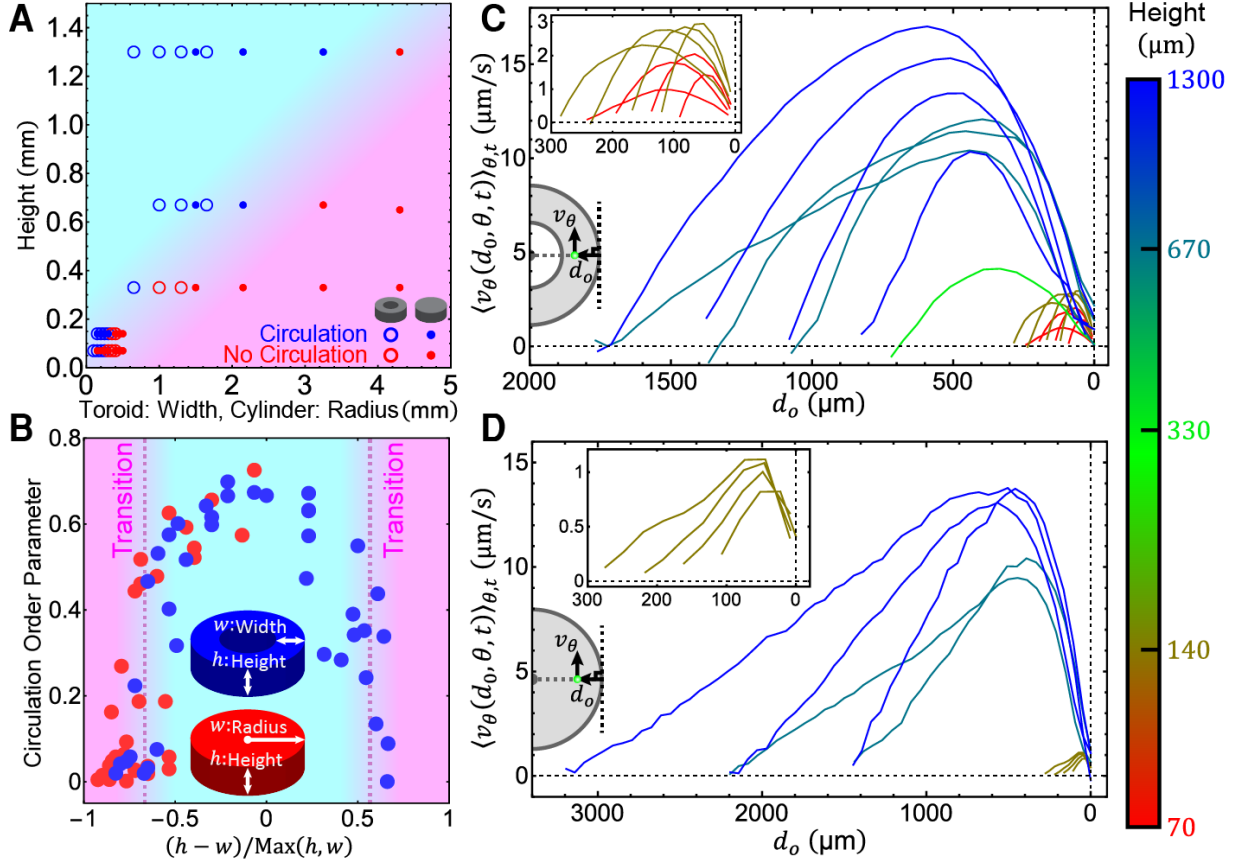

**Fig. 2. Phase diagram and velocity profiles of coherent flows in toroids and cylinders.** (**A**) A phase diagram indicating confinement geometries that support circular flows. The phase diagram is limited to thin confining geometries, $h \leq 2w$. Circles and dots represent toroidal and cylindrical confinements, respectively. (**B**) The circulation order parameter, $\Phi$, as a function of the confinement aspect ratio, $\alpha \equiv (h-w)/\text{Max}(h,w)$. $\alpha=0$ represents a square cross-section. Blue and red dots represent toroids and cylinders, respectively. For $\alpha<0.5$, $\Phi$ was measured at midplane. For $\alpha>0.5$, $\Phi$ was measured at a quarter plane due to the finite working distance of the microscope objective. (**C**) Flow velocity profiles in various toroidal confinements taken at a midplane. Outer diameters of toroids for blue, teal and green curves are 4,300 μm, and for olive and red curves, 2,000 μm. Inset: Flow profiles of toroids with width ≤ 300 μm. (**D**) Midplane flow profiles in various cylindrical confinements. Inset: Flow profiles in cylinders with radius ≤ 300 μm. The color bar on the right indicates the confinement heights.

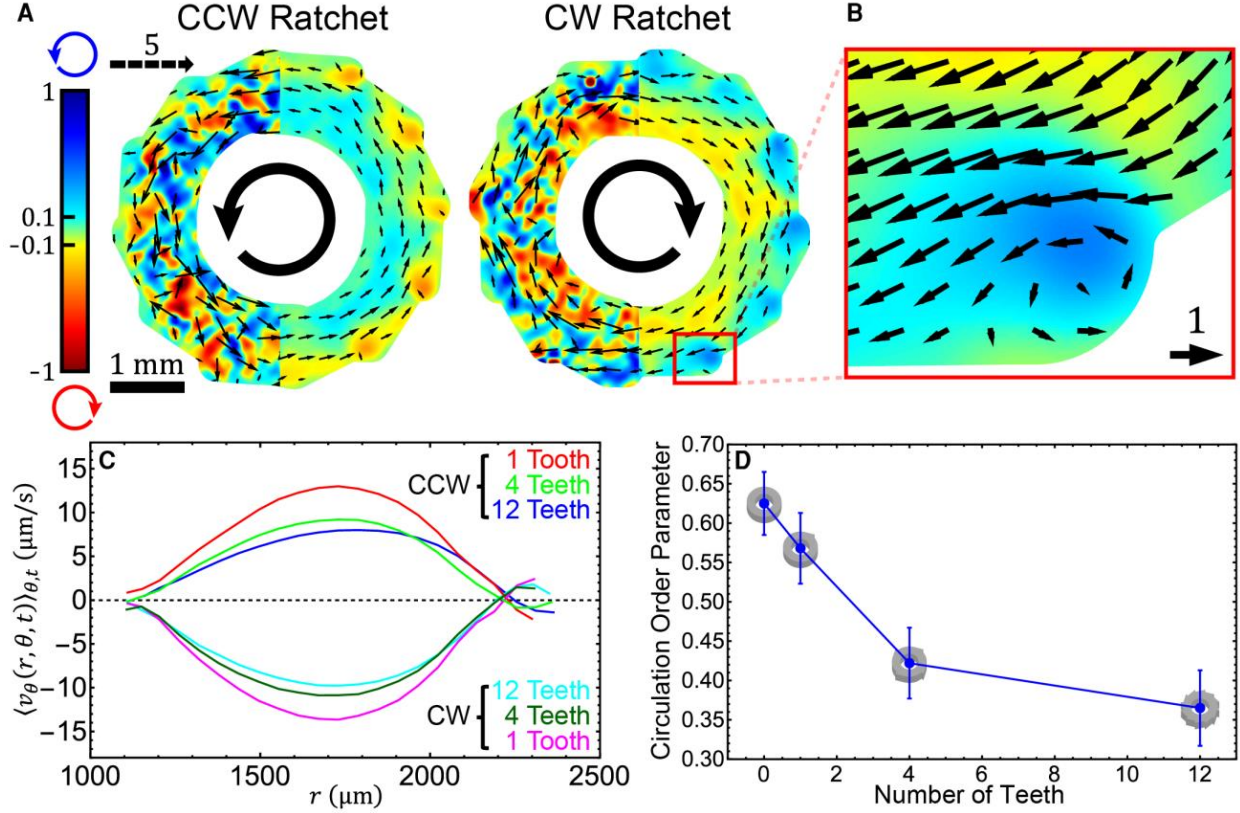

**Fig. 3. Controlling flow of coherent handedness.** (**A**) The handedness of the circular flows is controlled by decorating the toroid's outer surface with counter-clockwise or clockwise ratchets. The vector and color maps represent the local flow velocity and vorticity. (**B**) Velocity field and vorticity maps demonstrate that each notch induces a stationary vortex. (**C**) Midplane flow profiles in counter-clockwise and clockwise ratchets with varying notch numbers. Backflow from ratchets occurs near the outer edge, $r \approx 2{,}300$ μm. All channel heights, 1.3 mm. (**D**) The circulation order parameter as a function of the number of notches.

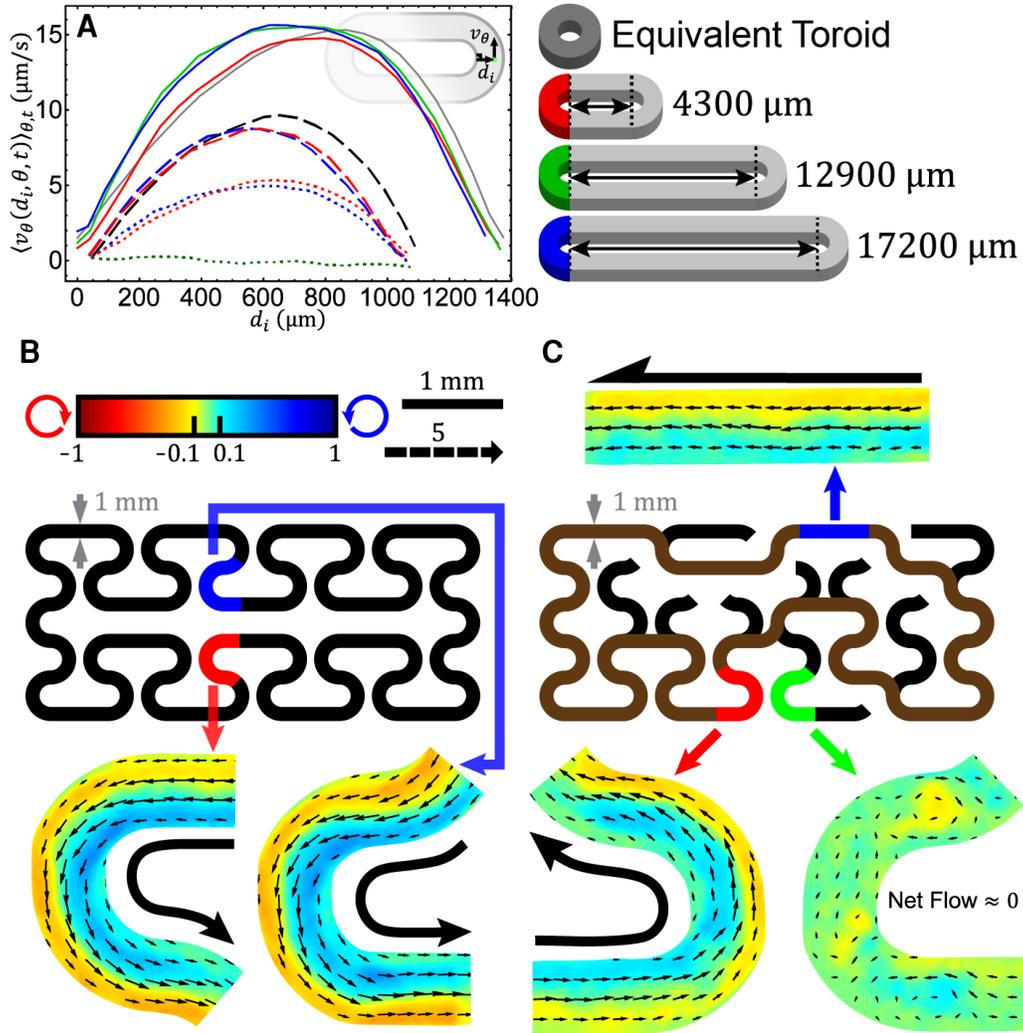

**Fig. 4. Circulating flows in closed tracks.** (**A**) Velocity profiles of coherent flows in racetrack geometries are independent of the length of the straight segment (solid curves). (**B**) Coherent flow in a 200-mm long racetrack. Due to limited observation windows we only observe the blue and red segments whose flow profiles are dashed same-color curves in A. The fluids in these segments flow as fast as in an equivalent toroid whose $w$=1 mm and $h$=1.3 mm (dashed black curve in A). (**C**) Self-organized flows can solve a 2D maze consisting of a closed loop (brown) and ten dead-end branches (black). We observe only the blue and red segments in the close loop and green segment in one branch. Their flow profiles are the dotted same-color curves in A. Dead-end branches inhibit net transport, while the closed loop supports a persistent circulating flow. All channel heights, 1.3 mm.

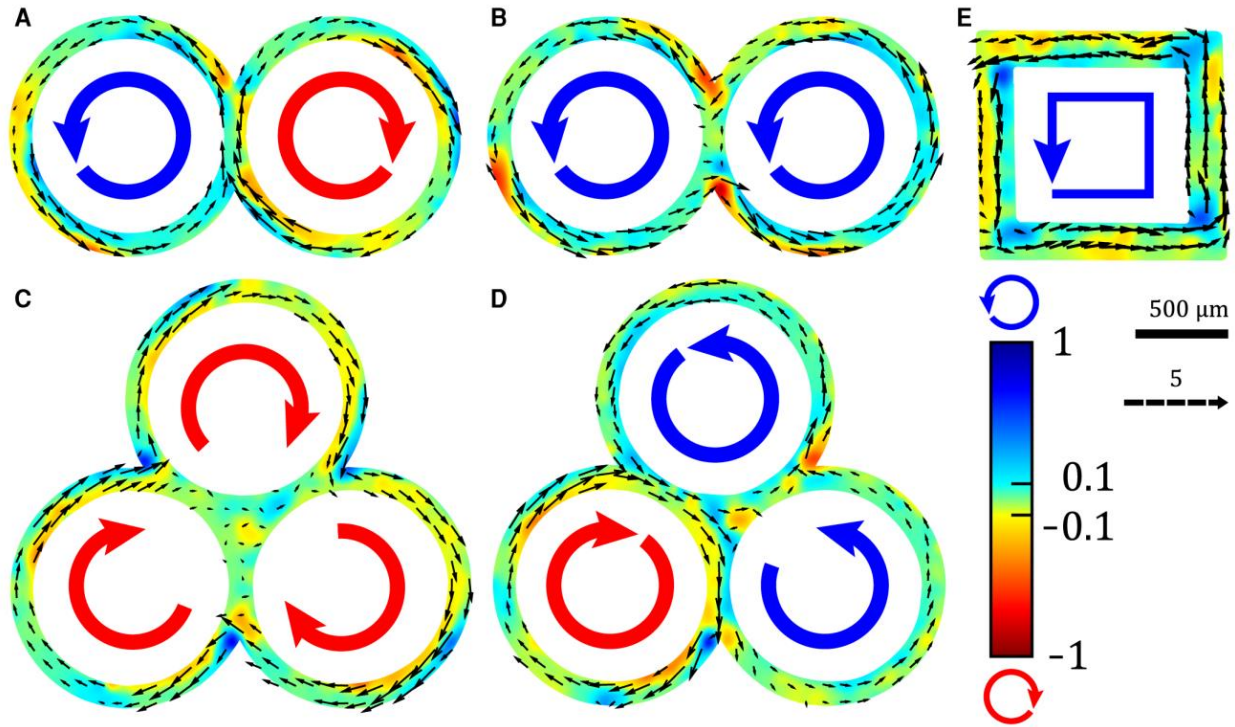

**Fig. 5. Coherent flows in miscellaneous confinements.** (**A-B**) Two distinct flow configurations in a pair of partially overlapping toroids. Confinement heights are ~60 μm. (**C-D**) Flow paths in three partially overlapping toroids. Confinement heights are ~60 μm. (**E**) Coherent flows in a square-like confinement. Channel height is ~75 μm.

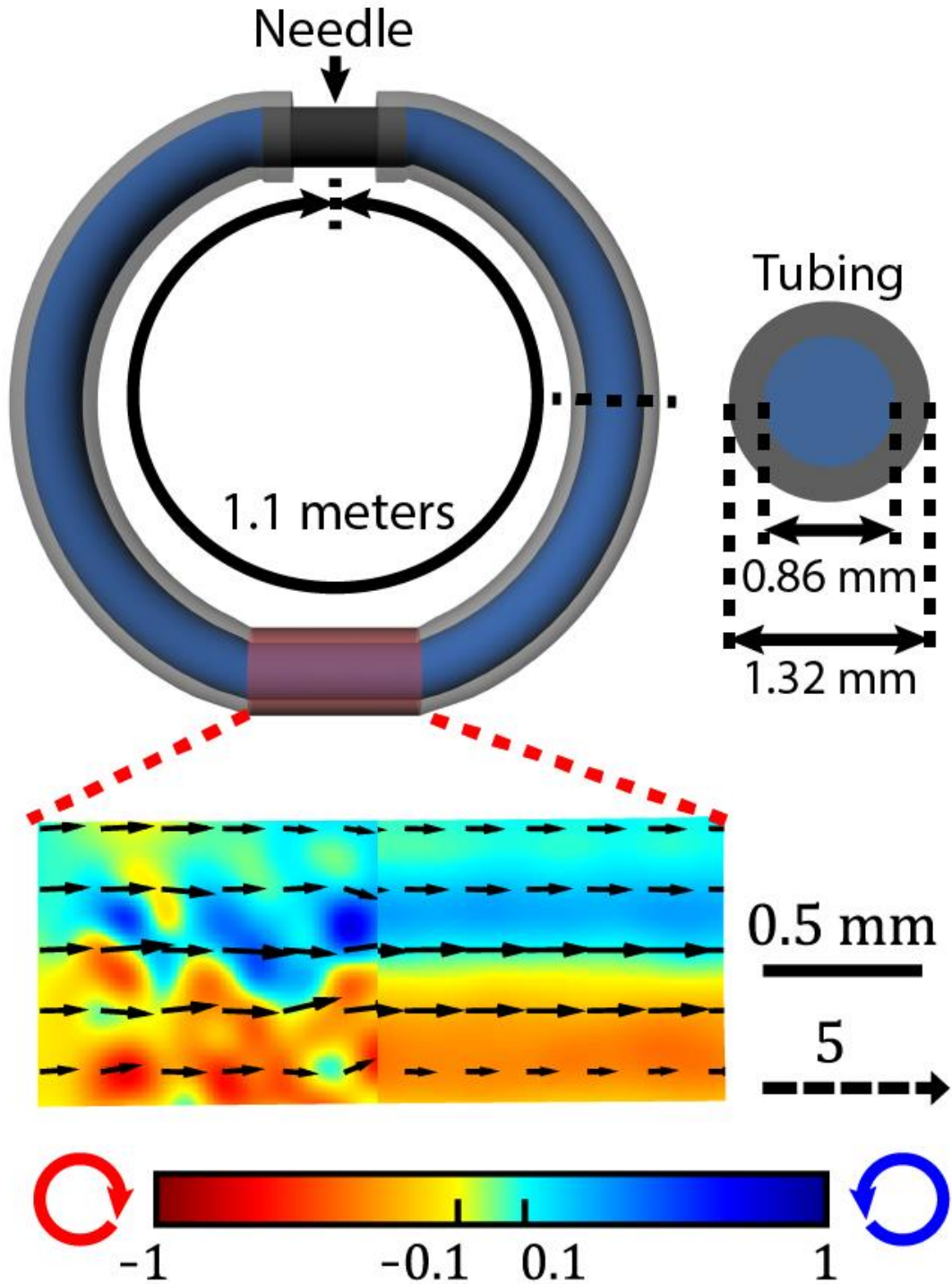

**Fig. 6. Coherent flows of an active fluid confined in meter-long tubes.** Red segment highlights the tube portion where fluid flows were observed and quantified.

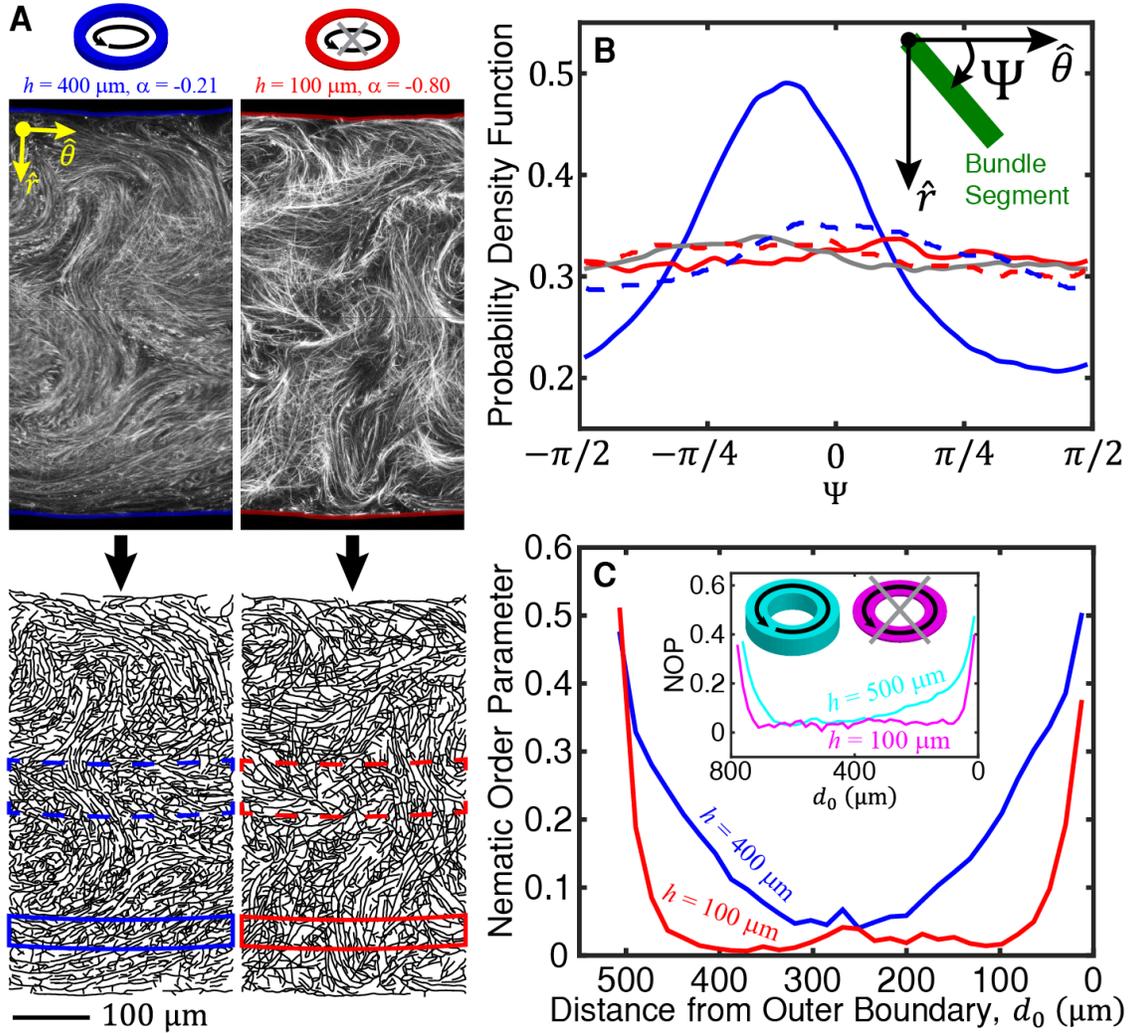

**Fig. 7. Transition to coherent flows correlates with the thickness of the nematic-wetting layer.** (**A**) (top) Fluorescence images of a microtubule network that drives the coherent and turbulent fluid flows taken at the channel midplane. (bottom) Reconstruction of the microtubule network that drives flow of active fluids in both the coherent and turbulent phase. (**B**) Orientational distribution functions (ODF) of microtubule bundle segments in the coherent (blue) and turbulent (red) state shown in A. The solid curves indicate that ODFs near the channel's outer boundary, while the dashed ones are measured at the channel center (solid and dashed boxes, respectively in A). Gray curve represents an isotropic suspension confined in a cuboid whose width = 18,000 μm, length = 14,000 μm, height = 90 μm. (**C**) Nematic order parameter (NOP) profile across the channel width. Inset: NOP profiles across 800-μm wide toroids in coherent (cyan) and turbulent (magenta) states.

**Movie 1, Coherent motion of toroidally confined active microtubules (red) and tracer particles (green):** Height and width of the toroid are 80 and 200 μm, respectively. To visualize flow in a higher resolution the tracer concentration is increased to ~0.01% v/v. Microtubules and tracers move coherently from top-right to bottom-left corners. Flow rates of microtubules and tracers reveal no measurable differences. Time stamp, hour : minute : second.

**Movie 2, Coherent and incoherent flows in toroids and cylinders:** Flows of active fluids in toroidal and cylindrical confinements that support and suppress coherent flows. Increasing confinement height induces a transition from incoherent to coherent flows. Heights of taller toroid (top-left) and cylinder (top-right) are 1.3 mm, and of shorter ones (bottom-left and -right) are 0.33 mm. Black vectors and color maps (left half in each panel) represent fluid velocity fields and vorticity distributions extracted from the motion of tracer beads (right half). Time stamp, hour : minute : second.

**Movie 3, Coherent flows in 12- and 1-tooth counter-clockwise (CCW) and clockwise (CW) ratchets:** Decorating the toroid outer surface with notches induced coherent flows in the direction of same handeness. Even one notch is sufficient to control flow direction. The notch sizes are ~300 μm. Top two ratchets (left: CCW, right: CW) have 12 teeth, and bottom ones (left: CCW, right: CW) have only 1 tooth. Heights are 1.3 mm. Time stamp, hour : minute : second.

**Movie 4, Coherent flows in concentric micron-sized toroids:** Toroids outer radii range from 175 to 1,525 μm. Varying the radii does not alter flow rates, implying coherent flow is independent of channel curvature. Tracks are shaded in blue and red for counter-clockwise and clockwise coherent flows. Channel heights are 60 μm. Time stamp, hour : minute : second.

**Movie 5, Coherent flows in racetracks with varying circumferences and an equivalent toroid:** Varying circumferences does not alter flow rate, implying coherent flow is independent of channel length. The lengths of straight paths from left to right are 17,200, 12,900, 4,300 and 0 μm, respectively. Heights and widths in the racetracks and toroid are both 1.3 mm. Time stamp, hour : minute : second.

**Movie 6, Circulations in two (doublet) and three (triplet) connected toroids:** Fluids in the doublet can develop a counter-clockwise and clockwise circulations (top-left) or two counter-clockwise circulations (top-right). In three adjoining toroids, fluids can flow in the outer edge

while leaving center quiescent (bottom-left) or develop two counter-clockwise and one clockwise circulations (bottom-right). Channel height is ~60 μm. Time stamp, hour : minute : second.

**Movie 7, Coherent (left) and incoherent (right) motion of active microtubule networks confined in toroids:** In a thin toroid (right, $h = 100$ μm) microtubules move irregularly. Increasing confinement height to $h = 400$ μm triggers a self-organized transition from turbulent to coherent flows (left). Time stamp, hour : minute : second.

**Movie 8, Demonstration of using SOAX to analyze three-dimensional bundle structure:** Magenta curves represent snakes which profile microtubule bundles; green dots represent their intersections. By using SOAX, bundle structures can be extracted from their confocal z-scanned images. Height, width and outer diameter of the toroid are 100, 200 and 2,000 μm, respectively.

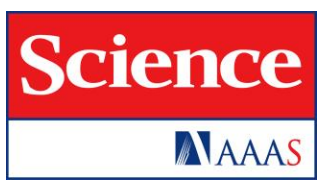

# Supplementary Materials for

Transition from turbulent to coherent flows in confined three-dimensional active fluids

Kun-Ta Wu, Jean Bernard Hishamunda, Daniel T.N. Chen, Stephen J. DeCamp, Ya-Wen Chang, Alberto Fernández-Nieves, Seth Fraden and Zvonimir Dogic

correspondence to: fraden@brandeis.edu, zdogic@brandeis.edu

**This PDF file includes:**

    Materials and Methods
    Figs. S1 to S5

**Other Supplementary Materials for this manuscript includes the following:**

**Materials and Methods**

The microtubule-based active fluids were confined within a microfluidic or millifluidic device sealed with clamps (Fig. S4).

Assembling kinesin clusters

Kinesin motors are dimers comprised of a 401-amino acid N-terminal domain, derived from *Drosophila melanogaster* kinesin labeled with a 6-his tag and a biotin tag. They were purified according to a previously published protocols (*48*). An isolated motor moves along a single microtubule protofilament towards its plus end. Linking kinesins into multimotor clusters enables them to bind multiple microtubules and drive interfilament sliding. The biotin-labeled motors were assembled into multimotor clusters using streptavidin tetramers. To assemble kinesin clusters, we mixed 1.5 µM kinesin motors with 1.8 µM streptavidin (Invitrogen, S-888) at a 1:1.2 ratio in microtubule buffer (M2B: 80 mM PIPES, pH 6.8, 1 mM EGTA, 2 mM $MgCl_2$) containing 120 µM dithiothreitol (DTT). The mixture was then incubated for 30 minutes at 4 °C, before being aliquoted and stored long-term at -80 °C.

Microtubule polymerization

Tubulin monomers were purified from bovine brains with two cycles of polymerizations and de-polymerizations in a high salt buffer (1 M PIPES buffer), followed with one cycle in M2B (*49*). To induce polymerization tubulin suspension (8 mg/mL in M2B) was mixed with 600 µM GMCPP (guanosine-5'-[($\alpha,\beta$)-methyleno]triphosphate, Jena Biosciences, NU-4056) and 1 mM DTT. For fluorescence imaging a 3% of tubulin monomers was labeled with Alexa-Fluor 647 (Invitrogen, A-20006). The tubulin mixture was then incubated for 30 min at 37 °C, followed by at room temperature for six hours to anneal. The resulting microtubules have an average length of ~1 µm (*24*). The polymerized microtubules were stored at -80 °C.

Microtubule-based active fluids

To assemble active fluids, 1.3 mg/mL microtubule suspension was mixed with kinesin-streptavidin motor clusters in a high-salt M2B (M2B + 3.9 mM $MgCl_2$) (*24*). Kinesin motors hydrolyze adenosine triphosphate (ATP) into an adenosine diphosphate (ADP) while stepping toward the microtubule plus end. To fuel such motion we added 1.4 mM ATP. To maintain constant ATP concentration and steady dynamics in active fluids, we have incorporated an ATP regeneration system comprised of 26 mM phosphoenol pyruvate (PEP) and 2.8% v/v stock pyruvate kinase/lactate dehydrogenase enzymes (PK/LDH) (Sigma, P-0294) (*50*). PK consumes PEP and converts ADP back to ATP, thus maintaining steady ATP concentration. To reduce photo-bleaching effects during fluorescence imaging, we added 2 mM trolox (Sigma, 238813) and an oxygen scavenging enzyme mixture comprised of 0.22 mg/mL glucose oxidase (Sigma, G2133), 0.038 mg/mL catalase (Sigma, C40), and 3.3 mg/mL glucose. To stabilize proteins we added 5.5 mM DTT as reducing agent. As the depletion agent we used 2% w/w Pluronic F127, which assembles into ~20 nm micelles (*51*). Finally, to track the flow induced by active fluid, we added ~0.0004% v/v Alexa 488-labeled 3-µm polystyrene particles, stabilized by Pluronic F127. Upon mixing all of the above ingredients non-equilibrium dynamics of active fluids can last for hours, until both ATP and PEP are depleted.

Engineering microfluidic chips

Millimeter-scale (≥0.3 mm) devices were manufactured by directly milling a 2-mm-thick COC plaque (TOPAS Advanced Polymers) with a CNC end-milling machine, which reads G-code translated from CAD with Cut2D (Vectric). Micron-scale devices were manufactured by embossing COC plaques (thermoplastics polymers) with patterned polydimethylsiloxane (PDMS, thermosetting polymers) slabs at ~180 °C (*52*). The slabs were molded by casting uncured PDMS resin (1:10 ratio of curing agent to base, Dow Corning) on a standard SU8 master and baked at 70 °C overnight. The master was photo-lithographed with photomasks printed from CAD. Millimeter- and micron-scale devices were subsequently cleaned with water by a ~20-min sonication. To protect proteins from direct surface contact, the COC chips were pre-coated with polymer brushes. Since COC is naturally hydrophobic we coated its surface with Pluronic by an overnight incubation in 2% w/w F127 in water.

Clamping

To confine active microtubules in a predesigned confinement, a patterned COC substrate was loaded with active microtubules, and subsequently covered with a 101-µm thick COC film (TOPAS Advanced Polymers). To prevent leakage or sample drying, the COC layers were sandwiched between a pair of aluminum frames held together by screws and bolts (fig. S6). A one-inch window was drilled in the middle of the aluminum frames to serve as an observation window. To prevent chips from falling through the windows, glass slides were placed across the windows. To ensure even pressure distribution, thin PDMS slabs were inserted between the COC chips and glass slides. After clamping, the samples remained sealed and hydrated for days. Microfluidic chips could be reused after being cleaned with ethanol and 20-minute sonication in water.

Analysis of fluid flows

Motion of active fluid flows was quantified by tracking the motion of tracer beads. To measure velocity fields, $\mathbf{v}(r,\theta,t)$, we analyzed sequential images of tracer beads with particle image velocimetry (PIV, PIVlab version 1.4) (*53*). Vorticity distribution is defined as $c(r,\theta,t) \equiv [\nabla \times \mathbf{v}]_z$. Velocity fields were normalized by their instant spatially averaged speed: $\mathbf{V}(r,\theta,t) \equiv \mathbf{v}(r,\theta,t)/\langle|\mathbf{v}(r,\theta,t)|\rangle_{r,\theta}$, and vorticity distributions were normalized by triple their instant standard deviations: $C(r,\theta,t) \equiv c(r,\theta,t)/[3\sigma(t)]$, where $\sigma(t)$ is the standard deviation of an instant vorticity distribution at time, $t$. To measure flow profile across a channel, tracers were tracked with Lagrangian particle tracking (*54*). Flow profiles were determined by averaging the θ-component of their velocities, $v_\theta$ over time ($t$), and orientations ($\hat{\boldsymbol{\theta}}$): $\langle v_\theta(r,\theta,t)\rangle_{\theta,t}$, while at quasi-steady states. To characterize flow coherency, we defined circulation order parameter, Φ as $\Phi \equiv \langle v_\theta/|\mathbf{v}|\rangle$, where $v_\theta$ is the θ-component of tracer velocity $\mathbf{v}$. Φ = 1 or -1 implies perfect counter-clockwise ($\hat{\boldsymbol{\theta}}$) or clockwise ($-\hat{\boldsymbol{\theta}}$) coherent flow, respectively. Φ = 0 implies either random flows or perfect radial ($\hat{\mathbf{r}}$) flows (not seen).

Elastomer confinements

To ensure that the development of coherent flows is not exclusive to COC-based microfluidic channels, we have confined active fluids in 3D tori inscribed in an elastomer (*55*). The elastomer consisted of 26% w/w 10-cst PDMS oil (Sigma-Aldrich, 378321-250ML) and 74% elastomer (Dow Corning 9041 silicone elastomer blend). To inscribe a torus, a 27-gauge blunt needle (SAI Infusion Technologies, B27-150) was inserted into the elastomer, followed by rotating the elastomer while pumping active fluids through the needle. The pumped fluid forms a torus after one full revolution. The radius of torus can be tuned by changing the distance between the needle tip and the elastomer rotation axis, while the torus thickness can be tuned by controlling the volume of the pumped fluid. Subsequently, the needle was extracted from the elastomer, yielding an isolated torus sealed within the elastomer. Active fluid contains Pluronic that coats the inner torus surface preventing microtubules adsorption. Unlike COC, the elastomer contains PDMS oil so the interface is fluid.

Tube confinements

To investigate broader applicability of our findings we loaded active fluids into a plastic tube (Scientific Commodities Inc., BB31695-PE/5, fig. S5B). The tube length was 1,060 mm; its outer and inner diameters were 1.32 and 0.86 mm, respectively. Tube ends were joined with a 50-mm long 20-gauge metallic needle (SAI infusion technologies, B20-150). The needle outer and inner diameters were 0.91 and 0.60 mm, respectively. To reduce evaporation from gaps between the needle and tube, vacuum grease was applied (Dow Corning). For imaging purpose the tube was fixed on a glass slide with tape. Circular profile of the tube distorts tracer images due to index mismatch. To reduce such distortions, the tube was immersed in uncured UV glue, straightening optical path from tracers to microscope objective.

Analyzing bundle structures

To correlate fluid flow with underlying active gel structure we z-scan microtubule networks with confocal microscopy when in coherent and turbulent states. From the midplane slice (thickness = 10 μm) we extract bundle structures which reveals orientations of bundle segments, $\Psi$, where $-\pi/2 \leq \Psi \leq \pi/2$ (56). The networks are monitored for one hour; from the sequential images we stack the extracted orientations. In an isotropic suspension the orientations are distributed more over ~0 and $\pi/2$ (blue curve in Fig. S5A). The distribution is shifted by ~45° if the images are rotated by -45°, indicating $\Psi$ is artificially biased along horizontal and vertical axes (red curve). To even such artificial bias we stack orientations extracted from images rotated by 0°, 30°, 60°, 90°, 120°, 150°, 180°, 210°, 240°, 270°, 300° and 330° (Fig. S5B). These stacked orientations are distributed not only more evenly (gray curve) but also disconnectedly from a similarly-made sample (black curve), indicating the artifacts are negligible and incomparable to noise.

From these orientations and their spatial distributions we can measure profiles of nematic order parameter (NOP), defined as NOP($d_0$) ≡ the largest eigenvalue of $Q_{ij}(d_0)$, where $Q_{ij}(d_0) \equiv \langle \hat{n}_i \hat{n}_j - \delta_{ij}/2 \rangle_{t,\theta}$, $\hat{\mathbf{n}} \equiv \cos[\Psi(d_0)]\hat{\boldsymbol{\theta}} + \sin[\Psi(d_0)]\hat{\mathbf{r}}$ is the director of a bundle segment at distance $d_0$ from outer boundary and $\delta_{ij}$ is Kronecker delta function. In either coherent or turbulent states NOP decays from a boundary with a decay length $l$, defined as NOP(distance $l$ from a boundary) ≡ NOP(at the boundary)/3 (Fig. 7C).

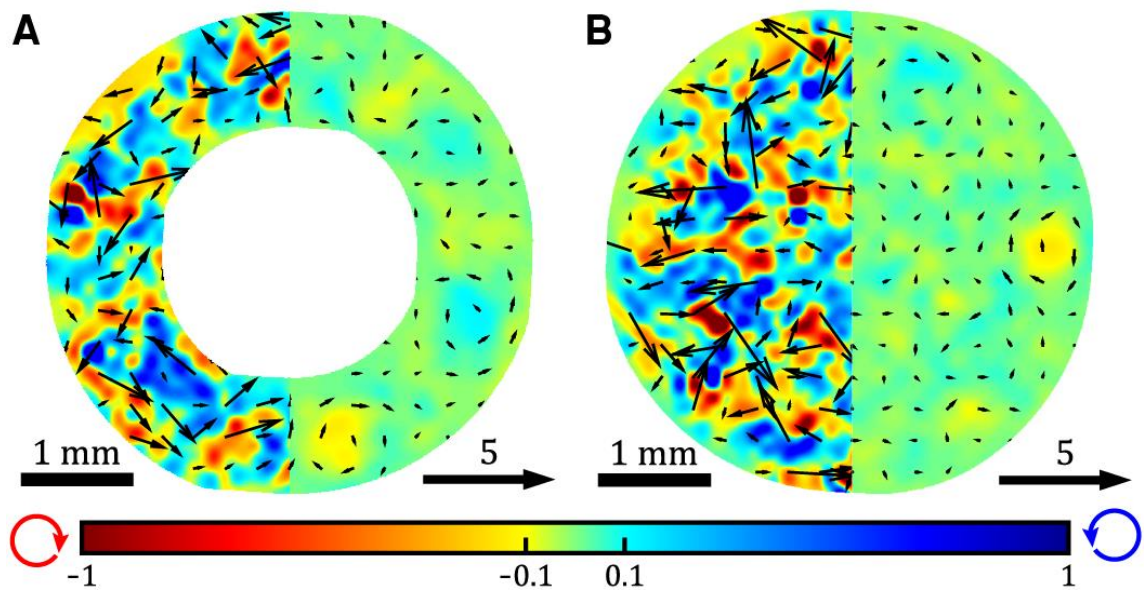

**Fig. S1.**
**Geometrical confinements with large aspect ratios suppress coherent flows.** (**A**) Incoherent flow in a toroidal geometry, with low circulation order parameter ($\Phi = 0.02$). Coherent flows were suppressed by reducing the height of the confining geometry from 1.3 to 0.33 mm (Fig. 1E). (**B**) Incoherent flow in cylinder with low circulation order parameter ($\Phi = 0.03$). To suppress coherent flows the height was reduced from 1.3 to 0.33 mm (Fig. 1F).

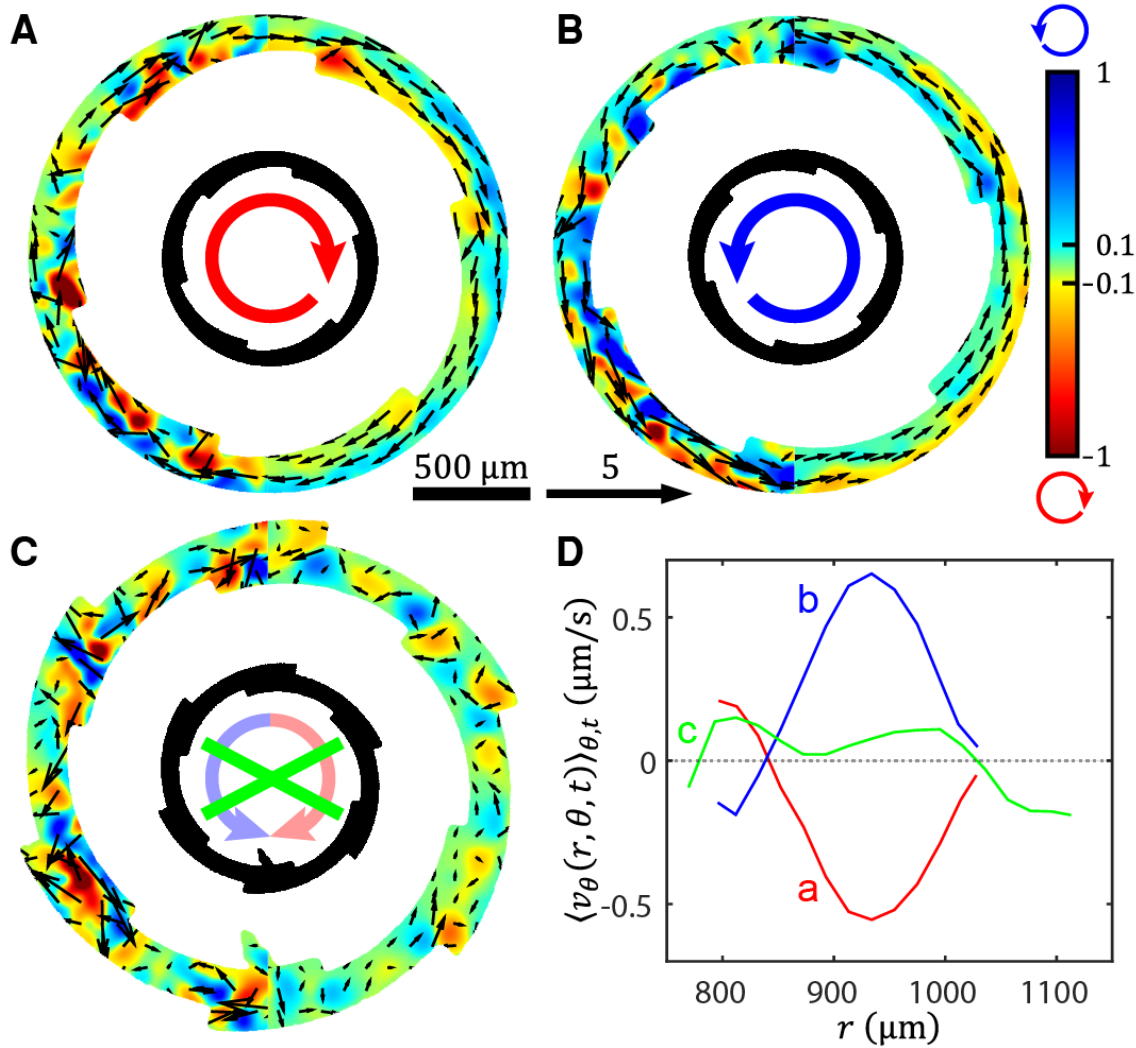

**Fig. S2**

**Ratchet geometries that suppress coherent flows.** (**A**) Clockwise circular flow directed by counter-clockwise saw-teeth decorating the inner toroid surface. Channel height is 70 μm. (**B**), Clockwise notches on the inner toroid surface direct counter-clockwise circulation. (**C**) Inner and outer saw-teeth patterns of the same handedness induce flows with the opposite handiness (Fig. 3A). Consequently, decorating both the inner and the outer surface with saw-tooth patterns suppresses circulating flows. (**D**) Velocity profiles of the self-organized flows in geometries shown in panels A-C.

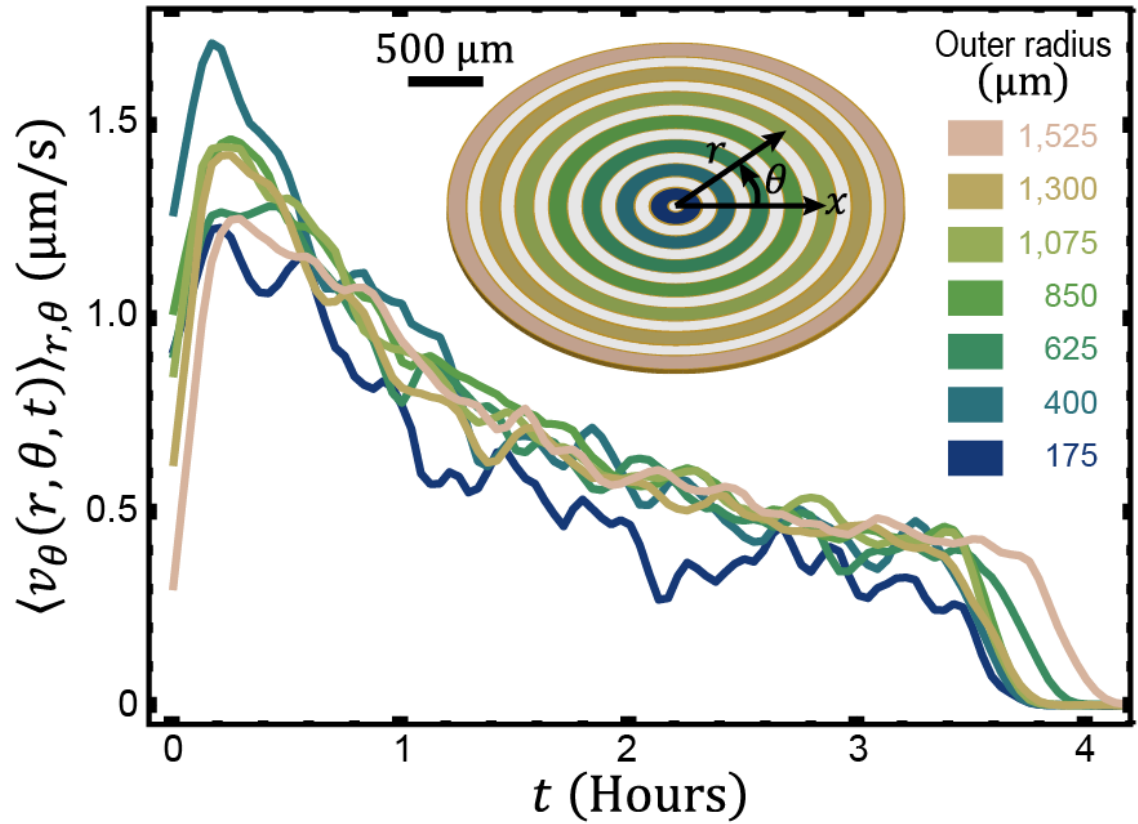

**Fig. S3**

**Flow profiles are independent of the toroid radius and curvature.** Evolution of the mean azimuthal velocities plotted for seven concentric toroids whose radii range from 175 to 1,525 µm. Flow velocities in toroids with different radii evolve similarly, demonstrating that the coherent flows are independent of the confinement curvature. Toroid heights are 60 µm.

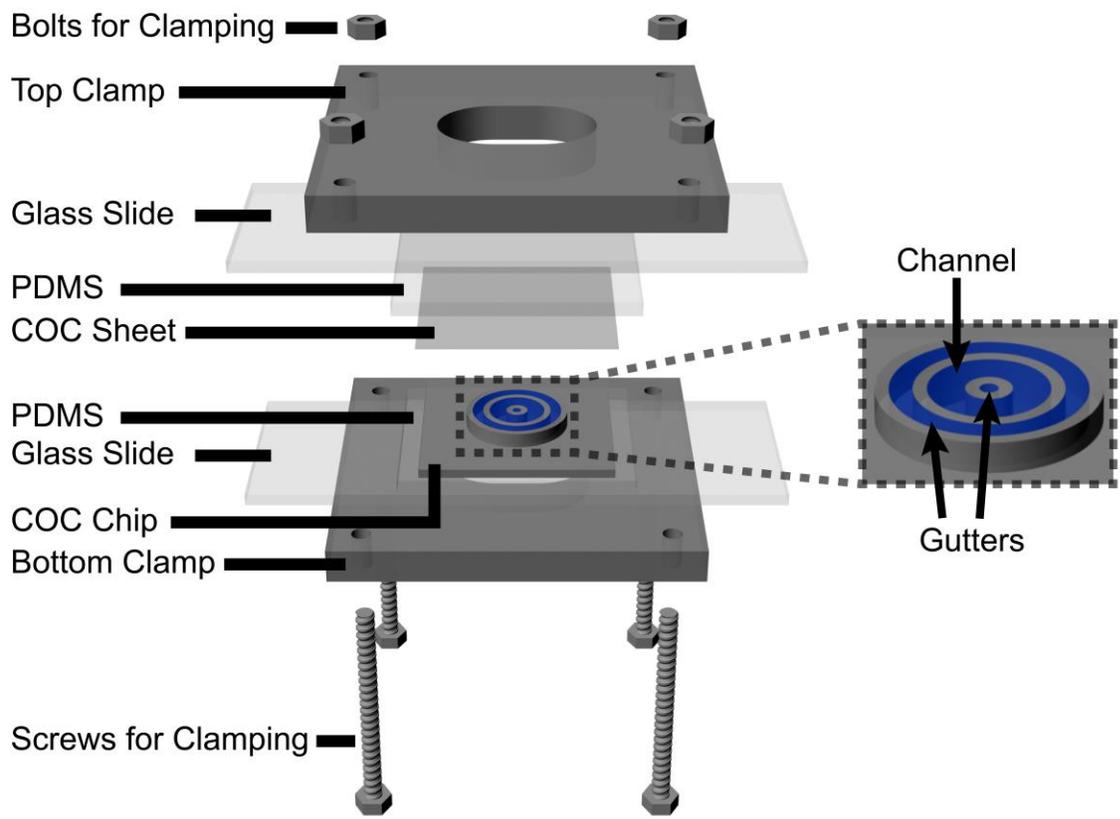

**Fig. S4**

**Clamping a microfluidic device for confining active microtubules.** To prevent evaporation the device is sealed with a clamp and the sample can remain hydrated for days.

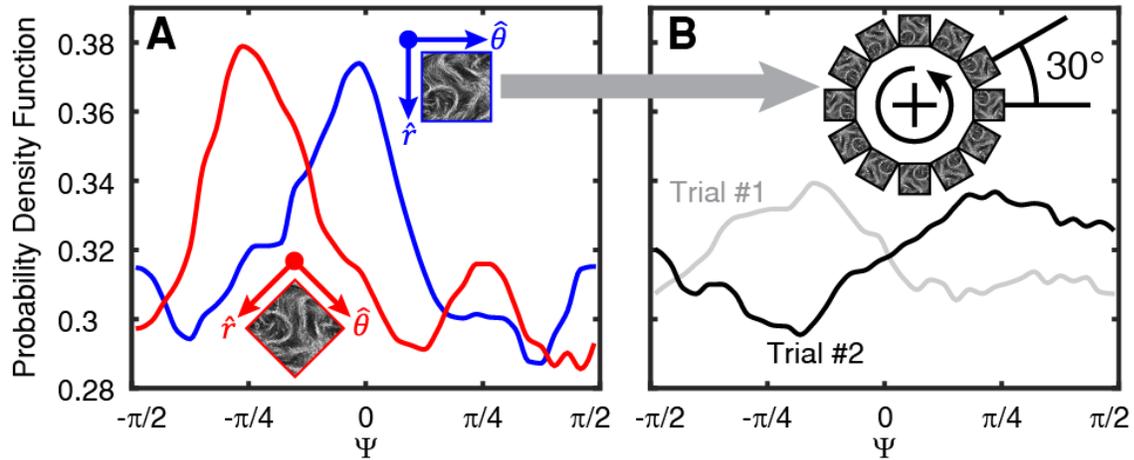

**Fig. S5**
**Orientational distribution of bundle segments in an isotropic suspension.** (**A**) The distribution from original (blue) and ($-\pi/4$)-rotated images (red). (**B**) The distributions of orientations stacked from one original and eleven rotated images. Gray and black curves represent two independent trials. The isotropic suspension is confined in a cuboid whose length, width and height are 18,000, 14,000 and 90 µm, respectively.